\documentclass[12pt,preprint]{aastex}

\begin{document}
\title{
UNIFIED MODEL OF RADIO AND X-RAY EMISSION OF NOVA CI CAM 1998
}

\author{N. N. Chugai}
\affil{Institute of Astronomy, RAS, Pyatnitskaya 48, 109017 Moscow, Russia}

\email{nchugai@inasan.ru}

\begin{abstract}

A unified model is proposed for the radio and X-ray 
outburst of nova CI Cam 1998 which suggests the shock interaction 
of nova shell with the circumstellar gas. The spherical model is 
able to describe kinematics of the radio shell 
together with the evolution of the 
radio and X-ray fluxes. However, the X-ray spectrum in this model is 
harder than the observed one. Better agreement with observations 
demonstrates the model in which the spherical shell interacts with 
the nonspherical circumstellar medium. The latter is made up of
the broad bipolar jets with the openning angle of $120^{\circ}$ and 
the dense equatorial wind. In the optimal model the kinetic energy of 
the nova shell is $\sim8\times10^{43}$ erg, while the shell mass lies 
in the range of $(1-5)\times10^{-7}~M_{\odot}$.

\end{abstract}

\section{Introduction}

Nova CI Cam was discovered 1998 March 31.64  
by the ASM detector of {\em RXTE}  as an X-ray outburst with the flux of 
139 mCrab in the range 2-12 keV \citep{Smi98,Rev99}.
At the earlier epoch March 31.36 this source was absent with the upper limit of 
40 mCrab. The extrapolation of the flux evolution backwards results 
in the outburst moment March 31.5, i.e., 50903.5 MJD; this is considered below 
as zero point of the nova shell expansion.
The X-ray source was identified with the binary of symbiotic type CI Cam 
\citep{WS98}, in which the luminosity is dominated by B-giant 
classified as B4 III-V[e] \citep{Bar06}. The binary has the orbital period 
of 19.4 d, eccentricity $e=0.62$ and major semiaxis 
$a=4.8\times10^{12}/\sin\,i$ cm, where $i$ is the inclination angle 
\citep{Bar06}. The optical outburst was 
observed starting from April 3 at the decline stage with the exponential time 
of 3.4 d \citep{Gar98,Cla00}.

Nova CI Cam 1998 turned out to be a bright source of the radio emission that 
was observed on several frequences starting from April 1.9 \citep{Hje98,MR04}.
The source mapping in radio band during several 
consequitive epochs staring from the beginning shows the source expansion with the 
velocity of $\sim12000$ km s$^{-1}$ assuming the distance of 5 kpc \citep{MR04}.
The radio emission is interpreted as a synchrotron 
emission of the shock wave propagating in the circumstellar medium (CSM). The 
radio evolution shows the strong early absorption at 2 GHz which can be caused 
by the synchrotron self-absorption (SSA) or/and $ff$-absorption in the 
CS gas \citep{MR04}. At the late stage $t\sim80$ d the radio 
shell shows an oval shape which is interpreted as a bipolar structure 
which can be interpreted 
 as a prolate spheroid, bipolar jets, or bipolar spherical clouds \citep{MR04}.
Interestingly, in this regard, \citep{Fil08} note that the spherical 
shock wave in the wind predicts significantly slower decay of the   
the X-ray emission compared with observations and that the better 
agreement shows the model of the shock wave in the equatorial disk.
 
The nature of the unseen stellar component of CI Cam and the origin of the 
outburst are unclear. There is a conjecture that nova CI Cam is the result 
of the outburst in the binary containing black hole or neutron star 
\citep{Fro98,Bel99}. On the other hand there are 
serious arguments that unseen component is the white dwarf (WD) 
and that the nova CI Cam 1998 is the result of the thermonuclear 
explosion on the surface of the white dwarf as in classical 
novae \citep{Ish04,Bar06,Fil08}.

Keeping in mind unusual nature of nova CI Cam it is desirable    
to advance in the interpretation of data on the radio and X-ray outbursts.
Particularly it would  be of interest to 
construct a model of the interaction of the nova shell with the CS gas 
that might account for the radio and X-ray data simultaneously. Such 
unified model 
possibly might provide us with interesting estimates of the crucial parameters,
e.g., energy and mass of the nova shell. The radio and X-ray data indicate 
the possibility of the nonspherical distribution of the CSM.
It would be sensible, therefore, to explore both spherical and nonspherical 
version of the model. This paper presents an attempt of the construction of 
the unified model in the framework of spherical and nonspherical approaches.
 
Model parameters essentially depend on the adopted distance ($d$) which 
is poorely known. Most popular are two options: $d=2$ kpc \citep{Cla00,Fil08}
and $d=5$ kpc \citep{MR04,Ish04}.
I will present results for both distance choices. 

\section{Dynamical model}

The basic model suggests the shock interaction of 
a spherical nova shell with a spherical CS medium. This model with some modification can 
be also used in the nonspherical case. Because the mechanism of the initial 
energy release is not completely clear, there is some uncertainty in the 
initial density and velocity distribution along the radius in the nova shell.
Two mechanisms of the shell ejection are conceivable: an explosion which 
suggests the energy release on time scale less or comparable to 
hydrodynamic 
time scale  and a relatively slow matter acceleration by the pressure gradient 
(optically thick wind). The latter is a major mechanism for the 
mass loss in classic novae \citep{Pri86,KH94}. In case 
of CI Cam this mechanism however cannot explain the high expansion velocity 
of the radio source which exceeds $5000$ km s$^{-1}$ for the distance $\geq2$ kpc 
\citep{MR04}. I therefore assume that the nova shell was 
ejected explosively. In this case one expects that the shell expands 
homologously soon after the passage of the shock wave, i.e., in the time of 
$\sim10^2$ s. However, this does not specify the density distribution.
Moreover, we do not know the matter distribution in the close vicinity of 
the white dwarf that could modify the initial density distribution of the 
nova shell.  

Bearing in mind above uncertainties it is reasonable to 
use a most simple dynamic model in order to minimize the number of 
free parameters;  the Sedov point explosion model is most appropriate. 
However one has to take into account also the initial deceleration phase.
Indeed, even for the low mass ejecta  $\sim3\times10^{-7}~M_{\odot}$ 
expanding in the dense CSM  $n\sim10^{10}$ cm$^{-3}$ the deceleration   
length scale is large $R\sim2\times10^{13}$ cm with the corresponding 
time scale of $\sim R/v\sim 1$ d. The deceleration phase is taken into account 
assuming the nova shell brakes as a whole with the energy conservation.
Equations of the energy and mass conservation is solved by Runge-Kutta forth 
order method.

Note, the formulated  deceleration problem for the shell of mass $M_0$ and the 
energy $E_0=(1/2)M_0v_0^2$ expanding in the stationary wind with the density
$\rho=w/4\pi r^2$ can be solved analitically
\begin{equation}
r=r_s\left[\left(\frac{t}{t_s}+1\right)^{2/3}-1\right]\,,
\label{eq-rsh}
\end{equation}
where $r_s=M_0/w$ is the radius at which the mass of the swept up shell 
becomes equal to 
the initial shell mass, and $t_s=(2/3)r_s/v_0$. At times $t\gg t_s$ this solution 
approaches Sedov solution $r\propto t^{2/3}$. For $M_0=3\times10^{-7}~M_{\odot}$,
and wind density parameter $w=10^{13}$ g cm$^{-1}$ the characteristic radius is 
$r_s=6\times10^{13}$ cm. For the initial velocity $v_0=6000$ km s$^{-1}$ the 
expansion enters 
the Sedov stage after about one day. The equation (\ref{eq-rsh}) 
is used to test numerical solution.

Following \citet{Fil08} I adopt the CSM distribution 
presented by a homogeneous core and an outer stationary wind. Specifically,
$\rho=\rho_0=const$ in the inner zone $r<r_0=2\times10^{13}$ cm and 
$\rho=\rho_0(r/r_0)^{-2}$ for $r>r_0$. The homogeneous core approximately 
takes into account the orbital motion of 
the less massive component with the velocity of $\approx 230$ km s$^{-1}$ which 
affects the slow wind flow with the velocity of $u=32$ km s$^{-1}$ 
\citep{Rob02} in the orbital zone $r\sim10^{13}$ cm. In fact, 
computations show that the shell dynamics only weakly depends on the 
variations of the density distribution in the region $r<r_0$.

A large volume of the parameter space was explored; below only two versions 
of the spherically-symmetric model are presented. They correpond to the minimal 
energy for which we are able to describe the expansion kinematics of the radio 
shell and to reproduce maximal X-ray flux for the adopted distance. 
The shell mass is poorely constrained because the expansion dynamics and 
X-ray ray emission are sensitive primarily to the energy and the CS density, 
not to the shell mass. In the model A for $d=2$ kpc and the energy 
$1.1\times10^{44}$ erg the shell mass lies in the range 
$(1-5)\times10^{-7}~M_{\odot}$ with the accepted optimal value of
$2\times10^{-7}~M_{\odot}$ defined by the radius of the radio source at the age 
$t\sim2$ d. In the model B for $d=5$ kpc with $E=1.6\times10^{45}$ erg
I adopt the same mass $2\times10^{-7}~M_{\odot}$. Models parameters are 
listed in Table 1 which, starting from the second column, shows the 
distance, shell mass and energy, CS density in the central zone, $r<r_0$, 
parameter of the density of the relativistic energy $\epsilon_r$,
spectral index of the relativistic electrons, and the filling 
factor of the radio-emitting clouds. The sense of some parameters is 
desribed in the appropriate place below.

The evolution of the shock wave radius for the models A and B 
(Fig. 1 and Table 1) 
satisfactorily reproduces the observed expansion of the radio source at the 
time scale of $\sim10^2$ d. The initial velocity in the model A is 
7400 km s$^{-1}$, somewhat larger than maximal velocities of novae shells 
but still within escape velocities $\approx1.1\times10^4$ km s$^{-1}$ for 
white dwarfs with masses  $\sim1.3~M_{\odot}$ \citep{Alt05}.
In the model B with $d=5$ kpc the initial velocity $2.8\times10^4$ km s$^{-1}$ 
is significantly larger than the velocity expected for any known type of 
cataclismic system with the white dwarf.

\section{X-ray emission of the shock wave}

In our model the reverse shock is absent. Fortunately, 
the role of the reverse shock in the X-ray emission is negligible: 
it is important only at the early deceleration 
stage $t\sim 1$ d. Moreover even at this stage it contributes only in the 
soft band $<1$ keV \citep{Fil09} because the speed of the reverse 
shock is relatively low. The X-ray emission of the forward shock is calculated 
assuming a homogeneous density of the shocked gas. The shock speed is assumed 
to be equal to the shell velocity found from the dynamics computation. The 
post-shock density is determined assuming strong shock limit. The 
contribution of the relativistic pressure (relativistic particles plus 
magnetic field) with a fraction $\eta$ of the total pressure is taken into account.
The compression factor in this case is $4+3\eta$ \citep{Che83}.
The inclusion of the relativistic pressure requires the itterative procedure: 
at the first step we compute dynamics and X-ray emission; afterwards  we calculate 
radio emission which provides us with the estimate of $\eta$ and then again 
recalculate X-ray emission. The electron temperatures is calculated from the 
equation of the heat exchange between ions and electrons $dT_e/dt=T_i/t_{eq}$,
where $t_{eq}=CT_e^{3/2}/n_i$ is equilibration time, $n_i$ is the ion concentration, 
$C$ is the factor specified in \citep{Spi62}. 
The initial temperatures in the shock $T_e$ and $T_i$ are taken  
according to Rankine-Hugoniot jump conditions while the age is adopted as an 
integration interval. The cooling function is taken according to 
\citet{SD93}. The X-ray flux density is presented by the 
spectrum $F(E)\propto E^{-0.5}\exp\,(-E/T)$ (where $E$ and $T$ are in keV).
Calculations show that always $T_e<T_i$. 

The model unabsorbed flux in the range of $3-20$ keV is shown in Fig. 2 
together with the unabsorbed flux derived from observations \citep{Fil08}. 
Insets show the evolution of the computed $T_e$ compared to the 
electron temperature recovered from the observed spectrum \citep{Fil08}.
The flux in the model A is consistent with observations only around maximum 
during the first day; afterwards the flux decays more slowly than the observed one.
The similar behavior shows the model flux in the paper \citep{Fil08}. 
The temperature significantly exceeds the observed one (Fig. 2) which reflects 
higher shock speed than is needed to account for the X-ray spectrum of 
CI Cam 1998. 
In the model B we face the same problem of the slow flux evolution as in model A, 
while the temperature shows even larger deviation from the observed one 
compared to model A. This difference is related with the substantially larger 
expansion velocity in the model B. Given disparity makes large distance 
($d>2$ kpc) highly unrealistic.

\section{Modelling radio emission}

The spectrum and high brightness temperature of the radio emission of 
nova CI Cam 1998 show 
that the radiation has a synchrotron origin, while the structure and kinematics 
of the radio source idicates that the emission originates from the shock wave 
propagating in the CS gas \citep{MR04}. 
The theory of the electron acceleration and generation of magnetic field in the 
shock wave cannot confidently predict parameters that define the radio emission.
I therefore use a parametric description. 

Below we assume that the magnetic field and relativistic electrons are 
homogeneously  distributed  on average behind the shock front within the 
layer of $h=0.2R$ where $R$ is the shock radius. The equipartition between 
nagnetic field and cosmic rays with the electron-to-proton ratio $K_{ep}=0.01$
is adopted. The ratio of the relativistic energy (magnetic field plus cosmic rays) 
to the internal energy in the shock wave ($\epsilon_r$) is a free 
pararmeter. The power spectrum of relativistic electrons in the energy range 
$E>E_{min}=0.5$ MeV is assumed $dN/dE\propto E^{-p}$, where spectral index 
$p$ is a free parameter. The model takes into account the SSA, 
the $ff$-absorption in the fully ionized wind with the temperature $10^4$~K; 
the Razin effect is also taken into account. The latter results in the 
exponential suppression of the spectrum at low frequencies  
$\nu<\nu_{\rm R}$, where $\nu_{\rm R}=(2/3)\nu_p^2/\nu_B$ ($\nu_p$ is 
the plasma frequency, $\nu_B$ is the gyrofrequency). The Razin effect can be 
taken into account using the flux prefactor \citep[cf.][]{Sim69}
\begin{equation}
G=C(\nu)\exp\,[-(3/2)^{1/2}\nu_{\rm R}/\nu]\,, 
\end{equation}
where $C(\nu)=1$ for $\nu\geq\nu_{\rm R}$ and 
$C(\nu)=(\nu/\nu_{\rm R})^{(3/2-p)}$ for  
$\nu<\nu_{\rm R}$. The modelling shows that the $ff$-absorption 
and Razin effect are of little importance compared to the SSA.

Preceding results, let us show that a homogeneous distribution 
of the radio-emitting material in the shell is inconsistent with the 
observational data. It is well known that the angular size or linear size
(if the distance is available) of the opaque radio source can be well 
determined from the flux \citep{Sli63}.
The size thus obtained  weakly depends on the magnetic field. Assuming the
equipartition one can exclude the magnetic field; moreover the size turns 
out to be insensitive to the ratio of magnetic field and relativistic 
electrons energy density \citep[cf.][]{Che98}. The dependence of the 
synchrotron luminosity at 2.25 GHz of the expanding source 
in the luminosity maximum on the peak time $t_m$ is shown in Fig. 3. This 
diagram is similar to that used for the analysis of radio supernovae 
\citep{Che98}.
The plot demonstrates that for the  shell with the homogeneous brightness the 
observed monochromatic luminosiy of CI Cam 1998 in the maximum on day 5 at 
the distance $2\leq d\leq5$ kpc corresponds to the expansion velocity 
 $700-1600$ km s$^{-1}$, significantly lower compared to the observed value
$\sim4000-6000$ km s$^{-1}$ at this epoch (Fig. 1a). This disparity means that 
the brightness distribution should be essentially inhomogeneous, i.e. the source has 
a cloudy structure. Note that the inhomogeneity of the radio brightness of 
CI Cam 1998 is apparent from the radio image \citep{MR04}.

The description of the inhomogeneous radio source requires additional 
parameters: filling factor ($f$) and the ratio of the cloud radius to 
the width of the radio-emitting shell ($\xi=a/h$). The intensity of 
the emergent radiation is then determined by the expressions
\begin{equation}
I_{\nu}=S_{\nu}[1-\exp\,(-\tau_{\rm eff})]\,, \qquad
\tau_{\rm eff}=\tau_{\rm g}[1-\exp\,(-\tau_{\rm c})]\,,
\label{eq-int}
\end{equation}
where $S=j_{\nu}/k_{\nu}$ is the source function, $\tau_{\rm g}=(3/4)(f/\xi)$ is 
the occultation optical depth (i.e., the average cloud number along the radius), 
$\tau_{\rm c}=(4/3)ak_{\nu}$ is the average optical depth of the cloud. 
The surface fraction covered by clouds (covering factor) is 
$1-\exp(-\tau_{\rm g})$. The expression (\ref{eq-int}) is used for the 
flux computation, which is sensible approximation.

The sensitivity of the radio flux behavior to the major parameters is 
shown in Fig. 4. The reference model A is characterized by 
$\epsilon_r=0.45$, $f=0.025$, and  $\xi=a/h=0.5$; three other cases show 
models in which each parameter is twice as smaller. The reduction of 
 $\epsilon_r$ decreases the flux at the optically thin stage and shifts 
the maximum towards the earlier epoch; the reduction of $f$ decreases the flux 
in all frequences; the reduction of $\xi$ shifts the maximum towards the 
earlier epoch since the cloud optical depth decreases. In the latter case the 
flux at the opaque stage is higher because the covering factor gets larger.
This plot shows how one can recover the optimal model via parameter variations.

The simulation results for the models A and B are shown in Fig. 5, while the 
parameter values are given in Table 1. In both models the relative cloud radius 
is assumed to be $\xi=0.5$ in order to secure maximal SSA for 
the moderate parameter $\epsilon_r$. Even with this prerequisite the required 
fraction of relativistic component is rather high $\epsilon_r=0.45$ and 
0.23 in models A and B respectively. As a result, the shock compression factor 
turnes out to be $>4$ so the fraction of the thermal pressure becomes 
notably lower 
than unity \citep{DV81,Che83}; this fact is taken into account 
for the calculation of the ion and electron temperatures. The filling factor 
$f\approx 0.025$ in both models while the covering factor $\sim0.1$ 
(front and rear sides of the shell are included). Models satisfactorily 
reproduce the flux evolution at different frequences, although at late stage 
the flux decreases more rapidly compared to the observations. Maximum at 
2.25 GHz is fully determined by the SSA whereas the optical depth due to 
$ff$-absorption is only 0.15 at $t=1$ d. The brightness temperature 
of radio-emitting 
clouds at 2.25 GHz in the flux maximum is $\sim2\times10^{10}$~K, 
markedly lower than the Compton limit $\sim10^{12}$~K.

The relativistic energy fraction $\epsilon_r$ in the preferred model A is 
rather high (0.45). This means that the kinetic energy in this model is 
close to the minimal permissible  value. On the other hand, the energy of the 
model A cannot be significantly larger because otherwise we would come to
the large X-ray luminosity. The recovered parameters of the 
model A therefore are close to the optimal values for the spherical models.

\section{Nonspherical model}

The bipolar structure of the CI Cam 1998 radio map at late times 
\citep{MR04} indicates that CSM might have two-component 
structure, e.g., rarefied medium in polar directions combined 
with the dense equatorial disk.
In this picture the radio emission could be associated with the fast shock waves 
running in the polar directions, while X-rays are  presumably emitted by
slow shocks propagating in the dense equatorial disk. \citet{Fil08} 
already proposed the equatorial disk for the interpretation of 
X-ray emission of CI Cam 1998. This model is preferred compared to the spherical 
one  because it is able to provide the fast decay of the X-ray flux. The ready  
scenario for the formation of the proposed structure of the CSM 
is absent. However, one could think that the combination of the 
slow wind with the velocity of 32 km s$^{-1}$ and of the fast disk wind with the 
velocity of $\sim10^3$ km s$^{-1}$ taken together with the gravitation perturbation 
from the unseen component with the orbital velocity of 230 km sÓ$^{-1}$ 
\citep{Bar06} might maintain the suggested structure of the CSM.

To simulate the radio and X-ray emission we consider the CSM composed of the 
rarefied bipolar outflow in cones with the opening angle $2\theta_0$ and 
dense equatorial disk with the polar angle range  $\theta_0<\theta<\pi-\theta_0$, 
where $\theta$ is counted form the polar axis $z$. The density in polar 
cones is constant ($\rho_0$) for $r\leq r_0=2\times10^{13}$ cm 
and $\rho=\rho_0(r_0/r)^2$ for $r>r_0$. In the equatorial disk the 
radial density distribution is similar but values are larger. The 
disk height over equatorial plane is assumed to be  $z_0\propto r$ for  $r<r_0$ and 
$z_0=r_0\cos \theta_0=constant$ for $r\gg r_0$. This behavior is approximated by 
the expression 
\begin{equation}
z_0=\frac{rr_0\cos \theta_0}{r_0+r}\,.
\end{equation}
We assume that the spherical nova shell interacts with the asymmetric 
CSM described above. The radio and X-ray emission is calculated separately for 
the polar cones (model Cp) and for the equatorial disk (model Ce) emploing the 
spherically-symmetric model. The total flux is a superposition of both components.
The resulting average electron temperature of X-ray-emitting plasma 
is defined as the flux weighted mean of 
temperatures of both components. It should be noted, however, that the 
contribution of X-rays from the bipolar component is small, so the flux and 
temperature is dominated by the disk shock. Assuming that the expansion 
of the radio source corresponds to the bipolar shocks we must admit that the 
inclination angle obeys to the condition $i>90^{\circ}-\theta_0$.

The nonspherical model describes the X-ray flux and the electron temperature 
(Fig. 6) significantly better compared to the spherical model A (Fig. 2). 
Parameters of the CSM for the bipolar cones and the disk are given in Table 1.
The X-ray emission is dominated by the disk shock, while the radio is 
related primarily with the bipolar cones. The relativistic energy fraction 
is large ($\epsilon_r=0.9$), so the model energy ($8\times10^{43}$ erg) is 
close to the minimal value; the initial velocity of the nova shell  
in the model is $6300$ km s$^{-1}$. Interestingly, the recovered energy is 
only 1.7 times larger than the energy of the spherical model in 
\citet{Fil08} for the case of the Sedov strong explosion models. 
 
Nonspherical model, alleviating the disparity between the model and observed 
rates of the flux decay, does not remove completely this contradiction. 
The similar contradiction is present in the disk model of 
\citet{Fil08}. The fast decay of the observed X-ray flux may be 
related with another effect being absent in the disk shock model, 
namely, the spherization of the post-shock flow due to outflow of shocked gas 
perpendicular to the equatorial plane. However, to take into account this 
effect requires the multi-dimensional hydrodynamics. 

The study of possible uncertainties implies that the nova energy and 
the density of the CSM are determined in our model with the accuracy of 
about 30\%. The mass 
of the nova shell is less certain because results weakly depend on the 
mass for the fixed value of energy. The optimal mass range 
$(1-5)\times10^{-7}~M_{\odot}$ is determined from the requirement of the 
description of the early ($t\approx2$) expansion.
Finaly, it should be stressed that the proposed nonspherical model is
illustrative; the actual distribution of the CSM may somewhat differ from the 
recovered optimal model.

\section{Discussion}

Our goal was the possibility to account for the radio and 
X-ray emission from nova CI Cam 1998 in the framework of a unified model of 
the nova shell interaction with the CSM. I show that the interaction of 
the spherical nova shell with the spherical CSM is able to describe the 
the evolution of the radio flux and the expansion kinematics of the radio source 
and only roughly to describe the X-ray emission assuming the distance of 2 kpc. 
The larger distance is less likely because it suggests the larger expansion velocity 
of the radio source and thus more harder X-ray spectrum in odd with observations.
For the distance of 2 kpc, however, more reasonable description can be found 
in the model of the interaction of the spherical shell with the nonspherical CSM. 
The latter is composed by the combination of the rarefied bibolar outflow and 
the dense equatorial disk.

The rarefied gas in bipolar cones can be associated with the fast bipolar wind 
characteristic of the binary systems with accreation disks. \citet{Rob02} 
report on the  detection in the $HST$ spectra taken in March 2000 of the 
ultraviolet 
lines of C {\sc iv} and Si {\sc iv} with P Cygni profiles and the expansion 
velocities of $\sim -1000$ km s$^{-1}$ in the blue absorption wing.
With this velocity and wind density from Table 1 
the mass loss rate related with the bipolar outflow is then
$\sim2.4\times10^{-6}~M_{\odot}$ yr$^{-1}$. For the equatorial wind velocity 
32 km s$^{-1}$ \citep{Rob02} the mass loss rate related with the 
equatorial wind turnes out to be $\sim0.5\times10^{-6}~M_{\odot}$ yr$^{-1}$.
The total mass loss rate is therefore $\sim3\times10^{-6}~M_{\odot}$ yr$^{-1}$, 
close to $\sim(1-2)\times10^{-6}~M_{\odot}$ yr$^{-1}$ 
obtained for the spherical model of X-ray emission of CI Cam 1998 \citep{Fil08}.

Given the openning angle of bipolar outflow $\sim120^{\circ}$ the 
inclination angle of the binary system should be $i>30^{\circ}$ to 
meet requirement that the shock tangential velocity should coinside with the 
physical expansion velocity. This inequality is consistent with another  
inequality $i>38^{\circ}$ found earlier from the requirement that the 
unseen component of the binary is the white dwarf \citep{Bar06}. 

Parameters of the optimal model of the nova CI Cam 1998
$E\sim 8\times10^{43}$ ÜÒÇ É $M\sim(1-5)\times10^{-7}~M_{\odot}$, bipolar 
structure, and the expansion velocity $\sim 6300$ km s$^{-1}$ are 
reminescent of the similar properties of recurrent nova RS Oph 2006.
Recently \citet{Rup08} studied bipolar structure of the radio 
source of RS Oph 2006
and found that the expansion velocity of bipolar shocks is 
$\sim 10^4$ km s$^{-1}$. They estimate the energy of bipolar outflow 
of RS Oph 2006 to be $\sim10^{44}$ erg, which is close to the energy of CI Cam 1998;
the estimated  ejecta mass  $\sim10^{-7}~M_{\odot}$ of RS Oph 2006  
is also close to the shell mass of CI Cam 1998.

Yet there are some apparent differences between these novae.
For comparable distances ($\sim2$) kpc and comparable extinction 
($A_V\sim 2.5$ mag) the optical flux in $V$ band at maximum of CI Cam 1998 
is $\sim40$ times lower compared to RS Oph 2006. On the other hand the X-ray flux 
of CI Cam 1998 in 0.5-10 keV band is $\sim20$ times larger, while radio 
flux at 5 GHz is $\sim7$ larger compared to RS Oph 2006. Binary systems are 
dissimilar as well. For CI Cam the binary period is 19 d and the donor is 
B-giant with the mass $\geq12~M_{\odot}$  \citep{Bar06}, 
whereas for RS Oph the binary period is 460 d and the donor is 
M-giant with the mass $\approx0.5~M_{\odot}$ \citep{DK94}. 
All these differences leave open the issue of the possible simiarity of the 
 outburst physics of CI Cam 1998 and RS Oph 2006.

{}

\newpage
%=====================================================================

\begin{table}
  \caption{Model parameters}
  \begin{tabular}{cccccccc}

\hline

Model & $d$    & $ M$           &   $E$         &    $\rho_0$            &                
        $\epsilon_r$ & $p$ & $f$  \\ 
     & kpc & $10^{-7}~M_{\odot}$ & $10^{44}$ erg & $10^{-15}$ g cm$^{-3}$ &
                      &      &  \\
      
\hline

 A  &   2   &     2           &     1.1      &       2           &
        0.45 & 2.3 & 0.025  \\
 B  &   5   &     2            &    16     &         3.2          &
       0.23 & 2.3 & 0.025 \\
Cp  &   2   &     2            &     0.8    &        0.6           &
        0.9  & 2.1 & 0.04 \\ 
Ce  &   2   &     2           &     0.8    &         4              &
       0.27  & 2.1 & 0.04 \\  
\hline
\end{tabular}
\label{t-mod} 
\end{table} 
%============================================================

\clearpage

%======================================================================
\begin{figure}
%\epsscale{.80}
\plotone{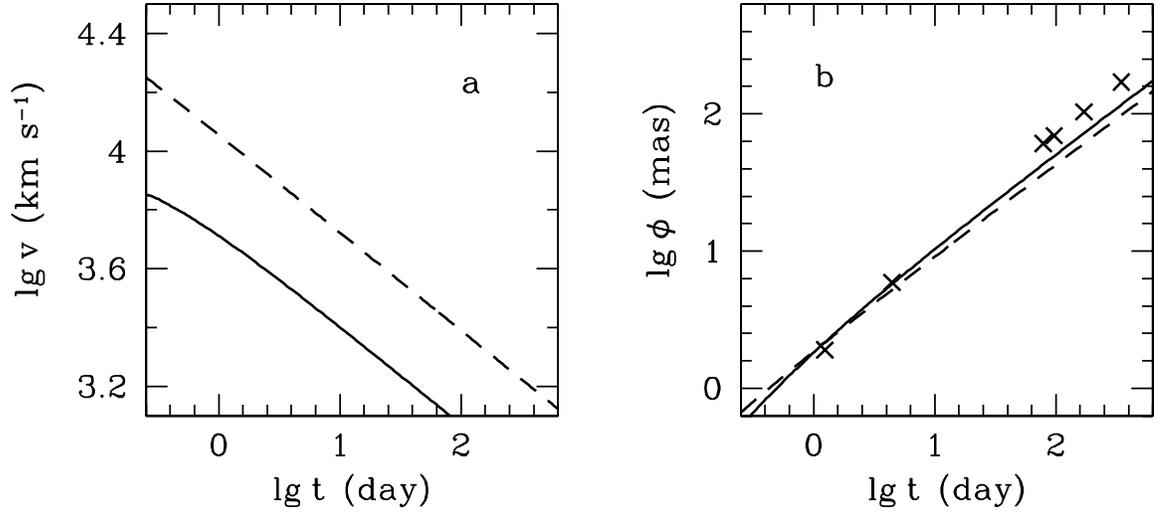}
\caption{
Evolution of the velocity and angular radius of the shock wave of CI Cam 1998 
in the model of the nova shell interaction with CSM. Left ({\em a}): 
the velocity in the model A ({\em solid line}) and in the model B ({\em dashed 
line}). Right ({\em b}): the angular radius in the model A ({\em solid line}) 
and in the model B ({\em dashed line}). Interferometric data ({\em crosses}) 
are taken from (Mioduszewski \&  Rupen 2004).
}
\end{figure}
%=====================================================================

\clearpage

%======================================================================
\begin{figure}
%\epsscale{.80}
\plotone{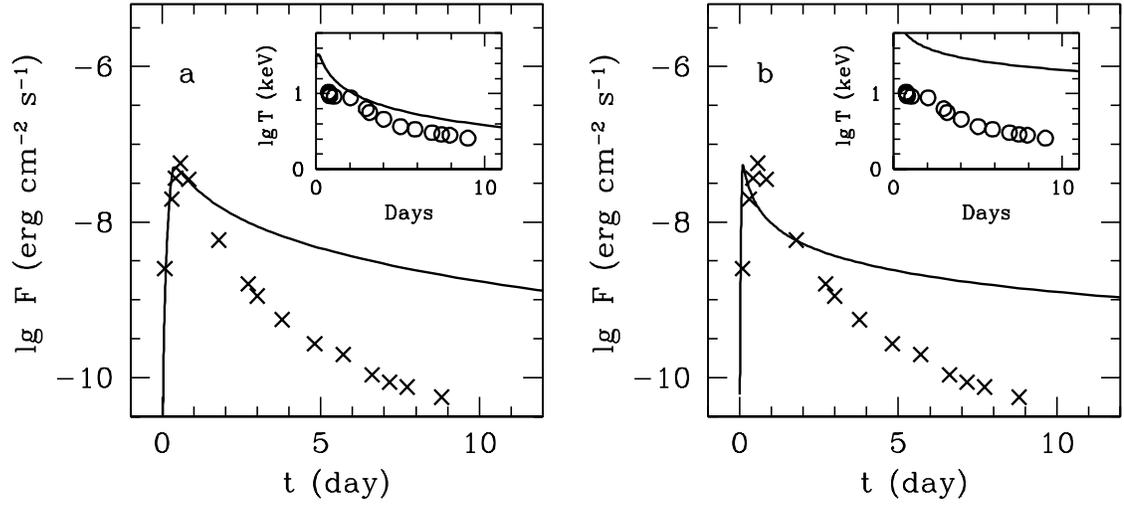}
\caption{
X-ray flux in 3-20 keV band in the model A ({\em a}) and model B ({\em b})
compared with the observational data (Filippova et al. 2008) ({\em crosses}).
Insets show electron temperature in the model ({\em line}) and observations 
(Filippova et al. 2008).
}
\end{figure}
%=====================================================================

\clearpage

%======================================================================
\begin{figure}
%\epsscale{.80}
\plotone{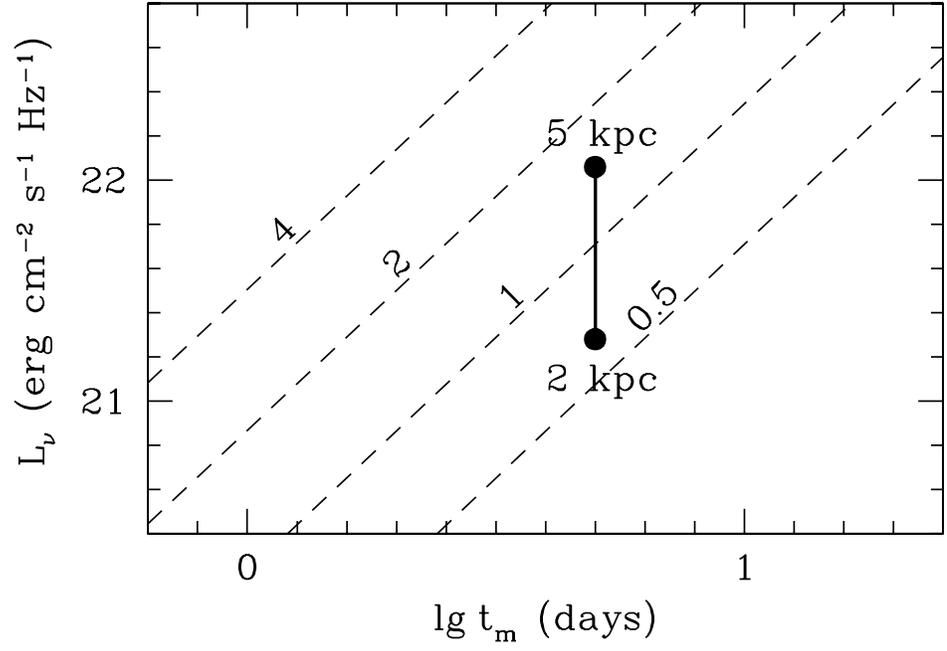}
\caption{
Peak synchrotron luminosity of the expanding spherical shell vs. time of the 
peak at the frequency 2 GHz. Numbers at the lines indicate the velocity in 
units of 1000 km s$^{-1}$.
}
\end{figure}
%=====================================================================

\clearpage

%======================================================================
\begin{figure}
%\epsscale{.80}
\plotone{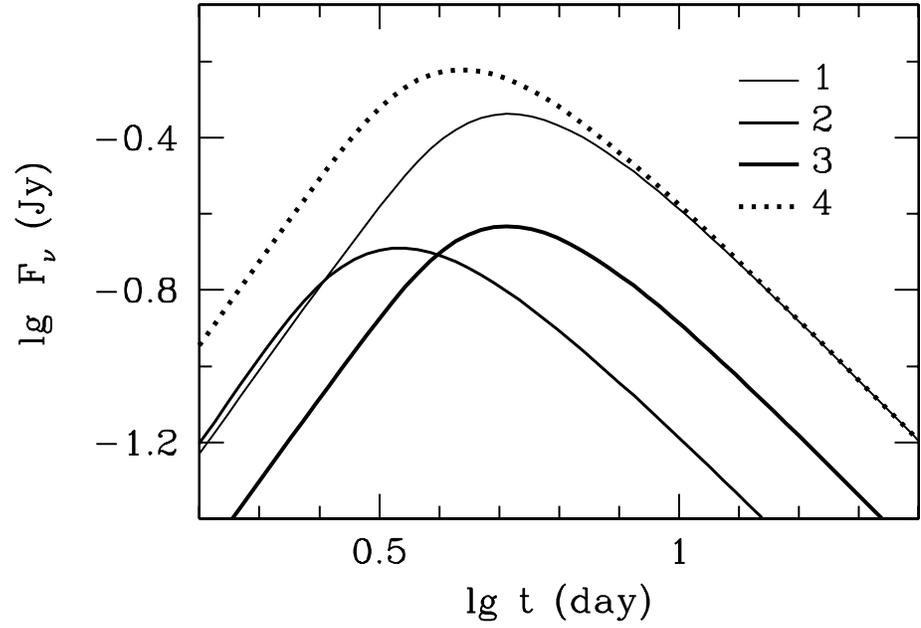}
\caption{
Flux evolution at 2 GHz for the expanding clumpy spherical shell. 
1 -- standard model (see text); 2 -- density of the relativistic component 
is twice as smaller; 3 -- filling factor is twice as smaller; 
4 -- cloud size is twice as smaller.
}
\end{figure}
%=====================================================================

\clearpage

%======================================================================
\begin{figure}
%\epsscale{.80}
\plotone{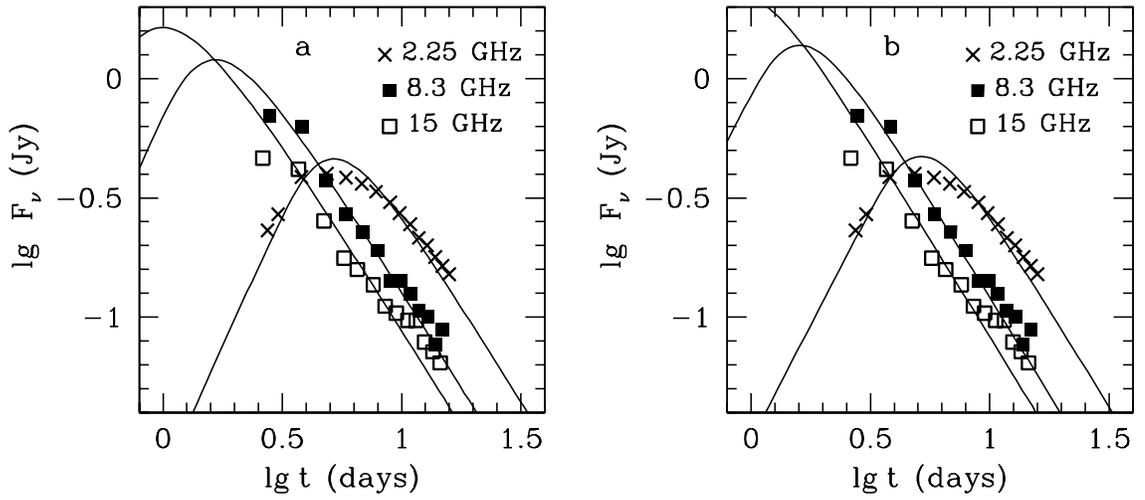}
\caption{
Flux evolution at frequences 2 GHz, 8 GHz, and 15 GHz in the model A 
({\em a}) and model B ({\em b}) compared to data (Clark et al. 2000).
}
\end{figure}
%=====================================================================

\clearpage
%======================================================================
\begin{figure}
%\epsscale{.80}
\plotone{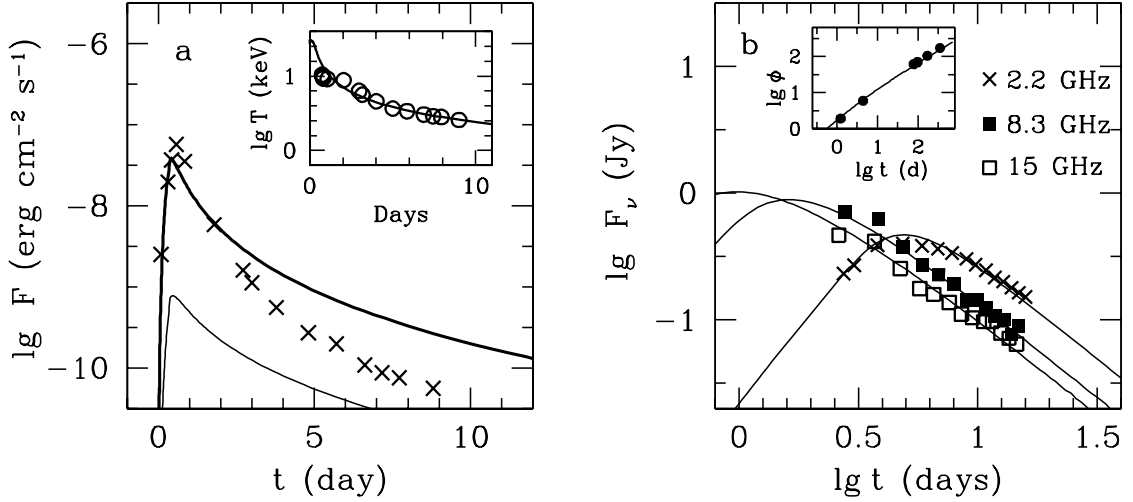}
\caption{
Evolution of X-ray and radio emission in the nonspherical model of CSM 
compared with data. Left ({\em a}): the same as in Fig. 2 but for 
nonspherical model; shown are the total flux ({\em thick line}) and 
contribution of shocks in polar cones ({\em thin line}). Right ({\em b}): 
the same as in Fig. 5 but for nonspherical model. The inset shows 
evolution of the angular radius (in mas) of the polar shock.
}
\end{figure}
%=====================================================================

\clearpage

\end{document}